Book review:

**Katy Börner,** *Atlas of Science: Visualizing What We Know***, Cambridge, MA/ London UK: The MIT Press, 2010; US$ 20.**

Katy Börner has written a wonderful book about visualization that makes our field of scientometrics accessible to much larger audiences. The book is to be read in relation to the ongoing series of exhibitions entitled "Places & Spaces: Mapping Science" currently touring the world. The book also provides the scholarly background to the exhibitions. It celebrates scientometrics as the discipline in the background that enables us to visualize the evolution of knowledge as the acumen of human civilization.

Katy Börner is deeply anchored in the tradition of the Enlightenment. In addition to Garfield and many of our other colleagues, Diderot and d'Alembert are prominently mentioned and indeed the book can be read as another Encyclopedia: one adapted for the internet age with its emphasis on access and visualization. As most of us know, the long-term project covers also the Networkbench (NWB) and the many other tools and platforms generated at the Cyberinfrastructure for Network Science (http://cns.iu.edu) of the School of Library and Information Sciences (Indiana University, Bloomington, IN). Open-source software is thus made available systematically in an integrated environment. This major effort and strong development has now been ongoing for more than seven years and the results pervade quantitative science studies, network analysis, and the various neighbouring fields.

The book's focus is wider than these scholarly disciplines and begins with Ptolemy and the other classics of building maps—of the earth, the heavens, and the sciences that study them. However, the focus shifts quickly to modern time and the wealth of illustrations—which are also made available online at http://scimaps.org/atlas/maps—makes the book attractive for the general audience. This is the only book in my library that I can read successfully with my granddaughter.

The introductory chapter is followed by two chapters entitled "The History of Science Maps" and "Toward a Science of Science." The aim is to be encompassing, ecumenical, and encyclopedic. The history of science mapping, in my opinion, rewrites this history from the perspective of the major efforts which the author, in collaboration with Kevin Boyack and Dick Klavans made for comprehensive science mapping during the last ten years (e.g., Boyack *et al*., 2006). Perhaps, this overshadows a bit the "prehistory" of science mapping such as, for example, the various mapping efforts of the 1980s when personal computing became available. The book is in this respect really a product of the 2000s: it mirrors the turn in our field from tables to visuals. The availability of a graphics environment since the mid-1990s changed this enterprise from a largely algorithmically driven one producing tables (e.g., Schubert *et al*., 1989) to one that strives for visualization. There is a certain tension between these two—for example, because a two-dimensional map reduces a multi-dimensional space—but Katy Börner demonstrates the extent to which surplus value is to be found in meeting the demands of larger audiences.

**"Building blocks"**

The book moves on by providing the "building blocks" (pp. 50 ff.) for a "science of science." This project of Bernal (1939), Price (1961), and others, was perhaps overambitious. It was partly dissolved by the reflexive turn in the history and philosophy of science, the sociology of scientific knowledge, actor-network theory, etc. Discourses carve and codify specific windows on the complexities under study, which may lead to "incommensurabilities" among paradigms (Kuhn, 1962) and therefore deep translation problems. These post-modernist insights are backgrounded in this book in favour of the grandiose vision of science as a human and cultural enterprise. The title is deliberately not "Atlas of the Sciences" in the plural, but "The Atlas of Science" in the singular! From this perspective, the deep divides and controversies can be appreciated as functional to the opportunity of "building bridges." This book is deliberately transdisciplinary and "Mode-2" (Gibbons *et al.*, 1994)

Science forecasting in terms of the performativity of science maps and scenarios for the future of science mapping follow in the fourth and fifth chapter. The various mapping opportunities are richly illustrated with materials from the exhibitions. David Zeller contributes with a chapter using a "hypothetical model of the evolution and structure of science" (at pp. 174 ff.). Both this chapter and the book transpire that scientific knowledge is a long-term and trans-generational enterprise which fascinates the author and to which she wishes to contribute by visualizing the visualizations. This reflexivity in action has given us a rich source at the unbelievably low price of less than US$ 20. Let

me advertise the book without any further hesitation: it is a celebration of our field of science.

Loet Leydesdorff

University of Amsterdam